\documentclass[12pt]{article}

\usepackage{amsmath}
\usepackage{amsfonts}
\usepackage{amssymb}
\usepackage{latexsym}
\usepackage{wasysym}
\usepackage{appendix}
\usepackage{setspace}
\usepackage{graphicx}		
\usepackage{subfigure}
\usepackage{epsfig}
\usepackage{rotating}

\begin{document}

\begin{flushright}
MAN/HEP/2011/14, FTUV-11-1909\\
{\tt arXiv:1109.3787 [hep-ph]}\\
September 2011
\end{flushright}

\bigskip

\begin{center}
{\bf \LARGE{On the Classification of Accidental Symmetries\\[3mm] 
                         of the Two Higgs Doublet Model Potential} }
\end{center}

\bigskip\bigskip

\begin{center}
{\large Apostolos Pilaftsis}~\footnote[1]{E-mail address: {\tt
apostolos.pilaftsis@manchester.ac.uk}}\\[3mm] 
{\it School of Physics
and Astronomy, LMS Consortium for Fundamental Physics, \\ University of
Manchester, Manchester M13 9PL, United Kingdom}\\[2mm]
{\it and}\\[2mm]
{\it Department of Theoretical Physics and
    IFIC, University of Valencia,\\ E-46100, Valencia, Spain}
\end{center}

\bigskip\bigskip\bigskip

\centerline{\bf ABSTRACT}
\vspace{2mm}

\noindent
Recently,     it     has     been    shown~\cite{BBP}     that     the
two-Higgs-doublet-model potential may exhibit a maximum of 13 distinct
accidental  symmetries.    Such  a   classification  is  based   on  a
six-dimensional bilinear scalar  field formalism realizing the SO(1,5)
symmetry group.   This note presents the  transformation relations for
each of the 13 symmetries in the original scalar field space and their
one-to-one correspondence  to the  space of scalar  bilinears, thereby
providing firm support for the completeness of the classification.


\thispagestyle{empty}

\newpage

There are  several theoretical and cosmological  reasons that motivate
us to go beyond the  Standard Model (SM) Higgs sector.  In particular,
the  so-called  Two  Higgs  Doublet  Model (2HDM),  where  the  SM  is
minimally  extended  with a  second  Higgs  doublet,  can provide  new
sources of  CP violation of  spontaneous~\cite{Lee:1973iz} or explicit
origin,     predict     stable      scalars     as     Dark     Matter
candidates~\cite{Silveira:1985rk},   and  give  rise   to  electroweak
baryogenesis~\cite{Kuzmin:1985mm} through  a strong first  order phase
transition~\cite{Cohen:1993nk}.  Unlike the SM, the 2HDM potential may
realize        a        large        number        of        different
symmetries~\cite{Deshpande:1977rw}, global or discrete, whose breaking
may  result  in  pseudo-Goldstone bosons~\cite{Weinberg:1972fn},  mass
hierarchies,  flavour-changing  neutral currents~\cite{Glashow:1976nt}
and  CP  violation~\cite{Branco:1980sz,Lavoura:1994fv,Botella:1994cs}.
The systematic analysis of the different possible symmetries and their
phenomenology    have    been    the    subject   of    many    recent
studies~\cite{Ginzburg:2004vp,Branco:2005em,Davidson:2005cw,Gunion:2005ja,Nishi:2006tg,
  Maniatis:2006fs,Ivanov:2006yq,Nishi:2007nh,Ivanov:2007de,Ferreira:2009wh}.

For given  choices of its  theoretical parameters, the  2HDM potential
may  exhibit three  different classes  of accidental  symmetries.  The
first  class of symmetries  pertains to  transformations of  two Higgs
doublets     $\phi_{1,2}$    only,     but    not     their    complex
conjugates~$\phi^*_{1,2}$, and are  therefore called Higgs Family (HF)
symmetries~\cite{Ginzburg:2004vp,Ferreira:2009wh}.       Known      HF
symmetries       include       the       $\mathrm{Z}_2$       discrete
symmetry~\cite{Glashow:1976nt},     the     Peccei--Quinn     symmetry
$\mathrm{U(1)}_{\rm  PQ}$~\cite{Peccei:1977hh}  and  the  HF  symmetry
$\mathrm{SU(2)}_{\rm HF}$~\cite{Deshpande:1977rw,Ivanov:2007de} acting
on the Higgs doublets $\phi_{1,2}$.

The second class of transformations relates the fields~$\phi_{1,2}$ to
their     CP-conjugates~$\phi^*_{1,2}$    and     are     termed    CP
symmetries~\cite{Ferreira:2009wh}. Known examples of this kind are the
CP1   symmetry  which   describes  the   standard   CP  transformation
$\phi_{1(2)}                                                        \to
\phi^*_{1(2)}$~\cite{Lee:1973iz,Deshpande:1977rw,Branco:1980sz},    the
CP2          symmetry          where         $\phi_{1(2)}          \to
(-)\phi^*_{2(1)}$~\cite{Davidson:2005cw}  and the  CP3  symmetry which
combines CP1  with an $\mathrm{SO(2)}_{\rm HF}$  transformation of the
fields $\phi_{1,2}$~\cite{Ivanov:2007de,Ferreira:2009wh}.

Nevertheless, there is a third class of symmetries which utilize mixed
HF and CP transformations that leave the ${\rm SU(2)_L}$ gauge kinetic
terms of $\phi_{1,2}$ canonical~\cite{BBP}.  Examples of this kind are
the O(8)  and ${\rm  O(4)\otimes O(4)}$ symmetries  in the  real field
space~\cite{Deshpande:1977rw}.  As  we will  show in this  note, these
mixed  HF/CP  transformations  play  an  important  role  to  properly
identify all the 13 accidental symmetries~\cite{BBP} that may occur in
the 2HDM potential. In particular,  based on the bilinear scalar field
formalism  realizing the  SO(1,5) symmetry  group, we  derive explicit
transformation relations for each of the 13 symmetries in the original
scalar  field space and  give their  one-to-one correspondence  to the
space of  scalar bilinears,  thereby proving the  self-consistency and
the completeness of the classification conducted in~\cite{BBP}.

To start with, let us write  down the general bare, local structure of
the  2HDM  potential~${\rm  V}$  in the  usual  doublet  field  space
$\phi_{1,2}$:
\begin{eqnarray}
  \label{eq:V2HDM}
\mathrm{V} \!& = &\! -\:\mu_1^2 (\phi_1^{\dagger} \phi_1) - \mu_2^2
(\phi_2^{\dagger} \phi_2) - m_{12}^2 (\phi_1^{\dagger} \phi_2) -
m_{12}^{*2}(\phi_2^{\dagger} \phi_1)
+ \lambda_1 (\phi_1^{\dagger} \phi_1)^2 + \lambda_2 (\phi_2^{\dagger} \phi_2)^2\nonumber\\
\!&&\! +\: \lambda_3 (\phi_1^{\dagger}
\phi_1)(\phi_2^{\dagger} \phi_2) + \lambda_4 (\phi_1^{\dagger}
\phi_2)(\phi_2^{\dagger} \phi_1)  + \frac{\lambda_5}{2} (\phi_1^{\dagger} \phi_2)^2 +
\frac{\lambda_5^{*}}{2} (\phi_2^{\dagger} \phi_1)^2\\\
\!&&\! +\: \lambda_6 (\phi_1^{\dagger} \phi_1) (\phi_1^{\dagger} \phi_2) + \lambda_6^{*}
(\phi_1^{\dagger} \phi_1)(\phi_2^{\dagger} \phi_1) + 
\lambda_7 (\phi_2^{\dagger} \phi_2) (\phi_1^{\dagger} \phi_2) +
\lambda_7^{*} (\phi_2^{\dagger} \phi_2) (\phi_2^{\dagger} \phi_1)\; .\nonumber
\end{eqnarray}
The potential  ${\rm V}$ contains  4 real mass  parameters, $\mu_1^2$,
$\mu_2^2$,  ${\rm Re}\, m_{12}^2$  and ${\rm  Im}\, m^2_{12}$,  and 10
real    quartic    couplings,    $\lambda_{1,2,3,4}$,   ${\rm    Re}\,
\lambda_{5,6,7}$ and ${\rm Im}\, \lambda_{5,6,7}$.

An alternative  formulation of the  2HDM potential may be  obtained by
introducing        the        8-dimensional        (8D)        complex
multiplet~\cite{BBP,Nishi:2011gc}
\begin{equation}
 \label{eq:Phi}
\Phi\ =\ \left( \begin {array}{c} \phi_1\\\noalign{\medskip}
 \phi_2\\\noalign{\medskip} i \sigma^2 \phi_1^*\\\noalign{\medskip} i
 \sigma^2 \phi_2^* \end {array} \right)\; ,
\end{equation}
where  $\sigma^2$   is  the  second  matrix  of   the  Pauli  matrices
$\sigma^{1,2,3}$.   Observe  that  each  doublet component  of  $\Phi$
transforms  covariantly under a  ${\rm SU(2)_L}$  gauge transformation
${\rm  U_L}$,   i.e.~$\Phi  \to  {\rm  U_L}\,   \Phi$.   Under  charge
conjugation,   the  multiplet   $\Phi$  satisfies   the  Majorana-type
property~\cite{BBP}:
\begin{equation}
 \label{eq:MajoranaPhi}
\Phi\ =\ \mathrm{C}\, \Phi^*\;,
\end{equation}
where $\mathrm{C} = \sigma^2  \otimes \sigma^0 \otimes \sigma^2$, with
${\rm C}  = {\rm C}^{-1} = {\rm  C}^*$ and $\sigma^0 =  {\bf 1}$ being
the $2\times 2$ identity matrix.

Given the $\Phi$-multiplet, we may go over to the bilinear field space
which   realizes  an   SO(1,5)   symmetry  group,   by  defining   the
{\em null} 6-vector~\cite{BBP}:
\begin{equation}
  \label{eq:RA}
\mathrm{R}^{\rm A}\ \equiv\ \Phi^\dagger\, \Sigma^{\rm A}\, \Phi\ =\ 
\left( \begin {array}{c} \phi_1^{\dagger}
 \phi_1+\phi_2^{\dagger} \phi_2\\\noalign{\medskip}\phi_1^{\dagger}
 \phi_2+\phi_2^{\dagger}
 \phi_1\\\noalign{\medskip}-i\left[\phi_1^{\dagger}
 \phi_2-\phi_2^{\dagger}
 \phi_1\right]\\\noalign{\medskip}\phi_1^{\dagger}
 \phi_1-\phi_2^{\dagger} \phi_2\\\noalign{\medskip}
 \phi_1^{\sf T} i \sigma^2 \phi_2 - \phi_2^{\dagger} i \sigma^2
 \phi_1^{*} \\\noalign{\medskip} -i \left[ \phi_1^{\sf T} i
 \sigma^2 \phi_2 + \phi_2^{\dagger} i \sigma^2 \phi_1^{*}
 \right] \end {array} \right) \; ,
\end{equation}
with   ${\rm  A}  =   0,1,2,3,4,5  =   \mu,4,5$.   The   six  $8\times
8$-dimensional matrices $\Sigma^{\rm A}$  may be expressed in terms of
double tensor products as follows:
\begin{eqnarray}
 \label{eq:sigmaA} 
\Sigma^{0,1,3} \!& = &\! \frac{1}{2}\, \sigma^0\otimes \sigma^{0,1,3}
\otimes \sigma^0\;, \qquad 
\Sigma^2\ = \ \frac{1}{2}\, \sigma^3 \otimes \sigma^2 \otimes
\sigma^0\; ,\nonumber\\ 
\Sigma^4 \!& = &\! -\ \frac{1}{2}\, \sigma^2 \otimes
\sigma^2 \otimes \sigma^0\;, \qquad 
\Sigma^5\ = \ -\ \frac{1}{2}\,
\sigma^1 \otimes \sigma^2 \otimes \sigma^0\; .
\end{eqnarray}
Note  that the  above six  matrices satisfy  the  Majorana condition:
$\mathrm{C}^{-1}         \Sigma^\mathrm{A}         \mathrm{C}        =
(\Sigma^\mathrm{A})^{\sf  T}$.    Further  details  of   the  matrices
$\Sigma^{\rm A}$ are given in~\cite{BBP}.

Having introduced the field-bilinear 6-vector $\mathrm{R}^\mathrm{A}$,
the potential ${\rm V}$ in \eqref{eq:V2HDM} can now be written down in
the quadratic form
\begin{equation}
  \label{eq:VR}
\mathrm{V}\ =\ -\ \frac{1}{2}\; \mathrm{M}_\mathrm{A}\,
\mathrm{R}^\mathrm{A}\: +\: \frac{1}{4}\; \mathrm{L}_{\mathrm{AB}}\,
\mathrm{R}^\mathrm{A} \mathrm{R}^\mathrm{B}\; ,
\end{equation}
where 
\begin{eqnarray}
\mathrm{M}_\mathrm{A} &=&  \left( \begin {array}{cccccc} \mu_1^2 +
 \mu_2^2\,,&2\mathrm{Re}(m_{12}^2)\,,&-2\mathrm{Im}(m_{12}^2)\,,&\mu_1^2 -
 \mu_2^2\,,& 0\,,& 0 \end {array}\right)\; , \\[3mm]
\mathrm{L}_{\rm AB} & = & \left(\! \begin {array}{cccccc} \lambda_1 +
 \lambda_2 + \lambda_3&\mathrm{Re}(\lambda_6 +
 \lambda_7)&-\mathrm{Im}(\lambda_6 + \lambda_7)&\lambda_1 - \lambda_2
 & 0 & ~0\\
\noalign{\medskip}\mathrm{Re}(\lambda_6 +
 \lambda_7)&\lambda_4+\mathrm{Re}(\lambda_5)&
 -\mathrm{Im}(\lambda_5)&\mathrm{Re}(\lambda_6 
 - \lambda_7)& 0 & ~0\\
\noalign{\medskip}-\mathrm{Im}(\lambda_6 +
\lambda_7)&-\mathrm{Im}(\lambda_5)&\lambda_4 -
\mathrm{Re}(\lambda_5)&-\mathrm{Im}(\lambda_6 -
\lambda_7)& 0 & ~0 \\
\noalign{\medskip}\lambda_1 - \lambda_2&\mathrm{Re}(\lambda_6 -
\lambda_7)&-\mathrm{Im}(\lambda_6 - 
\lambda_7)&\lambda_1+\lambda_2-\lambda_3 & 0 & ~0\\
\noalign{\medskip} 0 & 0 & 0 & 0 & 0 & ~0\\
\noalign{\medskip} 0 & 0 & 0 & 0 & 0 & ~0\end{array} \!\right)\; .
\end{eqnarray}
Evidently,  for  a ${\rm  U(1)_Y}$-invariant  2HDM  potential all  the
elements of  ${\rm M_A}$ and  ${\rm L_{AB}}$ involving  the components
${\rm A,B} =  4, 5$ vanish, where the  non-zero elements ${\rm M}_\mu$
and    ${\rm   L}_{\mu\nu}$    have    originally   been    calculated
in~\cite{Nishi:2006tg,Maniatis:2006fs,Ivanov:2006yq}.  As we will see,
however, this apparent redundancy  plays an important role to properly
identify  all accidental  symmetries that  may take  place in  a ${\rm
  U(1)_Y}$-invariant 2HDM potential.

We now consider ${\rm GL(8,\mathbb{C})}$ scalar-field transformations 
acting on the $\Phi$ multiplet that leave the SU(2)$_{\rm L}$ 
gauge-kinetic term of the Higgs doublets, $\frac{1}2\,(D_\mu \Phi)^\dagger 
(D^\mu\Phi)$, invariant, where $D_\mu = \sigma^0\otimes \sigma^0 \otimes 
(\sigma^0\,\partial_\mu + i g_w\, W^i_\mu \sigma^i/2)$ is the covariant 
derivative in the $\Phi$-space.  This restriction reduces the ${\rm 
GL(8,\mathbb{C})}$ transformations to unitary rotations ${\rm U} \in {\rm 
U}(4)$ in the $\Phi$-space, subject to the Majorana constraint~\cite{BBP}:  
${\rm U}^*\ =\ {\rm C}^{-1}\, {\rm U}\, {\rm C}$.  The latter condition 
implies that the generators $K^a$ of the Majorana-constrained U(4)  
group, denoted hereafter as ${\rm U_M}(4)$, should satisfy the important 
relation: \begin{equation}
  \label{eq:Kcond}
{\rm C}^{-1}\, K^a\, {\rm C}\ =\ -\; K^{a\,*}\ =\ -\; K^{a\,{\sf T}}\; .
\end{equation}
As  it can  be easily  checked, the  identity matrix  $\Sigma^0$ given
in~\eqref{eq:sigmaA} does not obey  the above condition, so $\Sigma^0$
cannot be one  of the generators of ${\rm  U_M}(4)$. Likewise, none of
the        other        5       matrices,        $\Sigma^{1,2,3,4,5}$,
satisfies~\eqref{eq:Kcond},  as  one obtains  the  opposite sign  (see
remark after~\eqref{eq:sigmaA}).   However, a careful  analysis yields
the following  10 generators $K^a$  (with $a = 0,2,\dots,9$)  of~${\rm
SU_M}(4)$:
\begin{eqnarray}
  \label{eq:Ka}
K^0 \!& = &\! \frac{1}{2}\, \sigma^3\otimes \sigma^0 \otimes \sigma^0\;, \qquad 
K^1\ = \ \frac{1}{2}\, \sigma^3 \otimes \sigma^1 \otimes \sigma^0\; ,\qquad
K^2\ = \ \frac{1}{2}\, \sigma^0 \otimes \sigma^2 \otimes \sigma^0\; ,\nonumber\\
K^3 \!& = &\! \frac{1}{2}\, \sigma^3\otimes \sigma^3 \otimes \sigma^0\;, \qquad 
K^4\ = \ \frac{1}{2}\, \sigma^1 \otimes \sigma^0 \otimes \sigma^0\; ,\qquad
K^5\ = \ \frac{1}{2}\, \sigma^1 \otimes \sigma^3 \otimes \sigma^0\; ,\nonumber\\
K^6 \!& = &\! \frac{1}{2}\, \sigma^2\otimes \sigma^0 \otimes \sigma^0\;, \qquad 
K^7\ = \ \frac{1}{2}\, \sigma^2 \otimes \sigma^3 \otimes \sigma^0\; ,\qquad
K^8\ = \ \frac{1}{2}\, \sigma^1 \otimes \sigma^1 \otimes \sigma^0\; ,\nonumber\\
K^9 \!& = &\! \frac{1}{2}\, \sigma^2 \otimes \sigma^1 \otimes \sigma^0\; .
\end{eqnarray}
Note that the generator $K^0$ is related to ${\rm U(1)_Y}$ hypercharge
rotations. Moreover, the 5 matrices  $\Sigma^{\rm I}$ (with ${\rm I} =
1,  2,  3, 4,  5$)  in~\eqref{eq:sigmaA}  and  the 10  matrices  $K^a$
in~\eqref{eq:Ka}  represent the  15~generators of  the  complete SU(4)
group.   In  particular,  because   of  the  different  even  and  odd
transformation properties  of $\Sigma^{\rm  I}$ and $K^a$  under ${\rm
  C}$ conjugation, the Lie commutators have the following structure:
\begin{equation}
  \label{faIJ}
\left[\, K^a\, , \, \Sigma^{\rm I}\, \right]\ =\ 2i\, f^{a{\rm IJ}}\,
\Sigma^{\rm J}\; ,
\end{equation}
where $f^{a{\rm  IJ}}$ is a subset  of the structure  constants of the
${\rm  SU}(4)$  group.   Note   that  the  Lie  commutators  involving
$\Sigma^{\rm I}$ or $K^a$ only close within themselves.

Since unitary transformations leave  the zero component ${\rm R}^0$ of
the 6-vector ${\rm R}^{\rm A}$  invariant, we only consider the action
of  ${\rm   SU_M}(4)$  on  its  `spatial'   components  ${\rm  R}^{\rm
I}$. Specifically, an  infinitesimal ${\rm SU_M}(4)$ transformation of
$\Phi$ changes ${\rm R}^{\rm I}$ by an amount
\begin{equation}
  \label{deltaRI}
\delta {\rm R}^{\rm I}\ =\ i\,\theta^a\; \Phi^\dagger\left[\,
  \Sigma^{\rm I}\, , \, K^a\, \right] \Phi\ =\ 
 2\theta^a\,  f^{a{\rm IJ}}\, {\rm R}^{\rm J}\; ,
\end{equation}
where $\theta^a$ are the  group parameters of ${\rm SU_M}(4)$. Observe
that   we   used~\eqref{faIJ}  to   arrive   at   the  last   equality
in~\eqref{deltaRI}.  The  corresponding  10  generators $T^a$  in  the
5-dimensional bilinear space ${\rm R}^{\rm I}$ may be calculated by
\begin{equation}
  \label{eq:Tadef}
(T^a)_{\rm IJ}\ =\ -i\, f^{a{\rm IJ}}\ =\ {\rm Tr} \left( \left[\,
  \Sigma^{\rm I}\, , \, K^a\, \right]\, \Sigma^{\rm J} \right)\; .
\end{equation}
In detail, we obtain 
\begin{eqnarray}
  \label{eq:Ta}
T^0 \!\!&=&\!\! 
\left(\!\begin{array}{ccccc}
0 & 0 & 0 & 0 & 0 \\
0 & 0 & 0 & 0 & 0 \\
0 & 0 & 0 & 0 & 0 \\
0 & 0 & 0 & 0 & i \\
0 & 0 & 0 &-i & 0 \\
\end{array}\!\right) ,\quad
T^1 = 
\left(\!\begin{array}{ccccc}
0 & 0 & 0 & 0 & 0 \\
0 & 0 &-i & 0 & 0 \\
0 & i & 0 & 0 & 0 \\
0 & 0 & 0 & 0 & 0 \\
0 & 0 & 0 & 0 & 0 \\
\end{array}\!\right) ,\quad
T^2 = 
\left(\!\begin{array}{ccccc}
0 & 0 & i & 0 & 0 \\
0 & 0 & 0 & 0 & 0 \\
-i& 0 & 0 & 0 & 0 \\
0 & 0 & 0 & 0 & 0 \\
0 & 0 & 0 & 0 & 0 \\
\end{array}\!\right) ,\nonumber\\[3mm]
T^3 \!\!&=&\!\! 
\left(\!\begin{array}{ccccc}
0 &-i & 0 & 0 & 0 \\
i & 0 & 0 & 0 & 0 \\
0 & 0 & 0 & 0 & 0 \\
0 & 0 & 0 & 0 & 0 \\
0 & 0 & 0 & 0 & 0 \\
\end{array}\!\right) ,\quad
T^4 = 
\left(\!\begin{array}{ccccc}
0 & 0 & 0 & 0 & 0 \\
0 & 0 & 0 &-i & 0 \\
0 & 0 & 0 & 0 & 0 \\
0 & i & 0 & 0 & 0 \\
0 & 0 & 0 & 0 & 0 \\
\end{array}\!\right) ,\quad
T^5 = 
\left(\!\begin{array}{ccccc}
0 & 0 & 0 & 0 & i \\
0 & 0 & 0 & 0 & 0 \\
0 & 0 & 0 & 0 & 0 \\
0 & 0 & 0 & 0 & 0 \\
-i& 0 & 0 & 0 & 0 \\
\end{array}\!\right) ,\nonumber\\[3mm]
T^6 \!\!&=&\!\! 
\left(\!\begin{array}{ccccc}
0 & 0 & 0 & 0 & 0 \\
0 & 0 & 0 & 0 & i \\
0 & 0 & 0 & 0 & 0 \\
0 & 0 & 0 & 0 & 0 \\
0 &-i & 0 & 0 & 0 \\
\end{array}\!\right) ,\quad
T^7 = 
\left(\!\begin{array}{ccccc}
0 & 0 & 0 & i & 0 \\
0 & 0 & 0 & 0 & 0 \\
0 & 0 & 0 & 0 & 0 \\
-i & 0 & 0 & 0 & 0 \\
0 & 0 & 0 & 0 & 0 \\
\end{array}\!\right) ,\quad
T^8 = 
\left(\!\begin{array}{ccccc}
0 & 0 & 0 & 0 & 0 \\
0 & 0 & 0 & 0 & 0 \\
0 & 0 & 0 & 0 &-i \\
0 & 0 & 0 & 0 & 0 \\
0 & 0 & i & 0 & 0 \\
\end{array}\!\right) ,\nonumber\\[3mm]
T^9 \!\!&=&\!\! 
\left(\!\begin{array}{ccccc}
0 & 0 & 0 & 0 & 0 \\
0 & 0 & 0 & 0 & 0 \\
0 & 0 & 0 &-i & 0 \\
0 & 0 & i & 0 & 0 \\
0 & 0 & 0 & 0 & 0 \\
\end{array}\!\right) .
\end{eqnarray}
These  are   exactly  the  10  generators  of   the  orthogonal  SO(5)
group. Consequently,  the relation~\eqref{eq:Tadef} represents  one of
the  central  results  of  this   note,  as  it  gives  an  one-to-one
correspondence between the generators  of ${\rm SU_M}(4)$ and those of
${\rm SO}(5)$.  Hence, we get the isomorphism: ${\rm SO(5)} \cong {\rm
  SU_M}(4)/{\rm  Z}_2$,  between the  $\Phi$-  and  the ${\rm  R}^{\rm
  I}$-space.   This  result  offers  firm  proof  of  the  equivalence
relation,  between  ${\rm   SU_M}(4)$  and  ${\rm  SO}(5)$,  presented
in~\cite{BBP}.

It is now obvious that  the maximal reparameterization group acting on
the  $\Phi$-space  in  the  2HDM  potential, which  leaves  the  ${\rm
  SU(2)_L}$ gauge kinetic term of $\Phi$ canonical, is
\begin{equation}
  \label{eq:Gmanifold}
{\rm G}^\Phi_{\rm 2HDM}\ =\ \left({\rm SU_M}(4)/{\rm  Z}_2\right) \otimes
{\rm SU(2)_L}\; . 
\end{equation}
The  group  ${\rm  G}^\Phi_{\rm  2HDM}$ includes  the  ${\rm  U(1)_Y}$
hypercharge group  through the generator $K^0$ of  ${\rm SU_M}(4)$, as
well as 9  other generators related to HF/CP  transformations.  On the
other hand, the ${\rm SU(2)_L}$ group generators may be represented as
$\sigma^0    \otimes   \sigma^0\otimes    (\sigma^{1,2,3}/2)$,   which
manifestly commute  with all generators of  ${\rm SU_M}(4)$.  Finally,
the quotient  factor ${\rm Z}_2$  appearing in~\eqref{eq:Gmanifold} is
needed to avoid double covering the group ${\rm G}^\Phi_{\rm 2HDM}$ in
the $\Phi$-space.

\begin{table}[!t]
{\small 
\begin{center}
\begin{tabular}{|r|c||ccccccccc|}
\hline
No & Symmetry & $\mu_1^2$ & $\mu_2^2$ & $m_{12}^2$ & $\lambda_1$
&  $\lambda_2$	& $\lambda_3$ & $\lambda_4$ &
${\rm Re}\,\lambda_5$ &	$\lambda_6 = \lambda_7$ \\
\hline
\hline

1 & $\rm Z_2 \times O(2)$	&	--	&	--
&	Real 
& -- &	-- & 
--	&	--	&
--	&	Real	\\ 

\hline

2 & $\rm (Z_2)^2\times SO(2)$	&
	--	&	--	&
0	& 
	--	&	--	&	--	&	-- 
&	--	&	0	\\  

\hline

3 & $\rm (Z_2)^3 \times O(2)$	&	--	&
$\mu_1^2$	& 
0	& 
	--	&	$\lambda_1$	&	--	&
--	&	--	& 0  \\ 

\hline

4 & $\rm O(2) \times O(2)$	&	--	&	--	&
0	&	--	&	--	&	--
&	--	&	0	&	0	\\  

\hline

5 & $\rm Z_2 \times [O(2)]^2$	&	--	&
$\mu_1^2$	& 
0	& 
	--	&	$\lambda_1$	&	--	&
--	&	$2\lambda_1 - \lambda_{34}$	 & 0	\\  

\hline  

6 & $\rm O(3) \times O(2)$	&	--	&
$\mu_1^2$ 
&	0	&	--	&	$\lambda_1$
&	--	&	$2\lambda_1 - \lambda_3$	&	0
&	0	\\ 

\hline

7 & $\rm SO(3)$ & --	& --	& Real & --	& -- & 	--& --&
$\lambda_4$ &	Real	\\    

\hline

8 & $\rm Z_2 \times O(3)$	&	--	&
$\mu_1^2$	& 
Real	& --	&	$\lambda_1$	& -- &	--	& $\lambda_4$
&	Real	\\   
\hline

9 & $\rm (Z_2)^2 \times SO(3)$	&	--	&	$\mu_1^2$
&	0	&	--	&	$\lambda_1$	& --	&
--	&	$\pm \lambda_4$	&	0	 \\   

\hline

10 & $\rm O(2) \times O(3)$	&	--	&
$\mu_1^2$	& 
0	&	-- &	$\lambda_1$	&	$2\lambda_1$	&
--	&	0	& 	0	\\  

\hline

11 & $\mathrm{SO(4)}$	&	--	&	--	&	0
& --	&	--	& --	&	0	&	0
&	0	\\  

\hline

12 & $\rm Z_2 \times O(4)$	&	--	&
$\mu_1^2$	& 0 
&	--	&	$\lambda_1$	& --	& 0	& 0 & 0	\\  

\hline

13 & SO(5)	&	--	&	$\mu_1^2$	&
0	& 
--	&	$\lambda_1$	&	$2\lambda_1$	&	0
&	0	&	0		\\  

\hline

\end{tabular} 
\end{center}  }
\caption{\it    Parameter   relations    for    the   13    accidental
  symmetries~\cite{BBP} related  to the ${\rm  U(1)_Y}$-invariant 2HDM
  potential  in  the  diagonally  reduced  basis,  where  ${\rm  Im}\,
  \lambda_5 =  0$ and $\lambda_6  = \lambda_7$.  A dash  signifies the
  absence of a constraint.  \label{tab:param} }
\end{table}

In order to  classify all possible HF/CP accidental  symmetries of the
2HDM potential, it is more  convenient to go over to the 5-dimensional
bilinear space ${\rm R}^{\rm I}$, where the maximal reparameterization
group  is ${\rm  G}^{\rm R}_{\rm  2HDM} =  {\rm SO(5)}$,  which leaves
${\rm  R}^0$  invariant.  Given  that  ${\rm  SO}(5)$  is the  maximal
symmetry  group  in   the  ${\rm  R}^{\rm  I}$-space,  Ref.~\cite{BBP}
classifies  all  possible   symmetries  derived  from  ${\rm  SO}(5)$,
including all its proper, improper and semi-simple subgroups.  Such an
analysis led  to a  maximum of 13  accidental symmetries for  the 2HDM
potential,  which  are presented  in  Table~\ref{tab:param}. The  same
table shows the  parameter restrictions for each of  the 13 symmetries
in a specific bilinear basis~\cite{Gunion:2005ja}, where ${\rm L}_{\rm
  IJ}$  is made  diagonal  by  an ${\rm  SO(3)}  \subset {\rm  SO(5)}$
rotation~\cite{Maniatis:2011qu}.   In this  diagonally  reduced basis,
one has the restrictions:
\begin{equation}
  \label{eq:reduced}
{\rm Im}\, \lambda_5\ =\ 0\;,\qquad \lambda_6\ =\ \lambda_7\; ,
\end{equation}
thus reducing to 7 the number of independent quartic couplings for the
2HDM potential.   From Table~\ref{tab:param},  we observe that  all 13
symmetries include  ${\rm SO(2)} \cong  {\rm U (1)_Y}$ as  a subgroup.
Note that the  parameter relations pertinent to the  13 symmetries are
chosen, so as to manifestly lead to CP-invariant scalar potentials.

It is worth commenting that only two discrete factors, $({\rm Z}_2)^2$
and  $({\rm  Z}_2)^4$,  are  allowed,  as being  the  only  admissible
subgroups of SO(5),  where ${\rm Z_2}$ is the  reflection group of one
of the components ${\rm R}^{\rm I}$.  More explicitly, the standard CP
(or CP1) discrete  symmetry may be represented as  $\Delta_{\rm CP1} =
{\rm  C}  =  \sigma^2   \otimes  \sigma^0  \otimes  \sigma^2$  in  the
$\Phi$-space, and  the usual discrete `${\rm Z}_2$'  (CP2) symmetry as
$\Delta_{\rm  Z_2}  =  \sigma^0  \otimes  \sigma^3  \otimes  \sigma^0$
($\Delta_{\rm CP2} = \sigma^2  \otimes \sigma^2 \otimes \sigma^0$). In
the  ${\rm  R}^{\rm I}$-space,  the  transformation  matrices (or  the
generating group elements) associated  with the CP1, `${\rm Z}_2$' and
CP2 discrete symmetries are respectively given by
\begin{eqnarray}
  \label{eq:D}
{\rm D}_{\rm CP1} \! & = &\! {\rm diag}\,(1,-1,1,1,-1 )\; ,\qquad
 {\rm D}_{\rm Z_2}\ =\ {\rm diag}\,(-1,-1,1,-1,-1 )\; ,\nonumber\\
{\rm D}_{\rm CP2} \! & = &\! {\rm diag}\,(-1,-1,-1,1,-1 )\; .
\end{eqnarray}
As a consequence, both the traditional `${\rm Z}_2$' symmetry and CP2
are actually isomorphic to the $({\rm Z}_2)^4$~symmetry.

It  is straightforward  to identify  the generators  pertinent  to the
continuous HF/CP  symmetries of the  2HDM potential in  the diagonally
reduced  basis~\eqref{eq:reduced}.  Specifically,  the 2HDM  potential
possesses a continuous symmetry, {\em iff}
\begin{equation}
\left[\, T^a\, ,\, {\bf L}\, \right]\ =\ 0\;, \qquad 
T^a\,{\bf M}\ =\ 0\; ,
\end{equation}
where  ${\bf L}$ and  ${\bf M}$  denote the  $5\times 5$  matrix ${\rm
  L}_{\rm IJ}$ and  the 5-dimensional vector ${\rm M}_{\rm  I}$ in the
reduced  basis,  respectively.   Given the  one-to-one  correspondence
between $T^a$ and  $K^a$ generators, it is not  difficult to determine
the transformation relations associated  with a given continuous HF/CP
symmetry in the $\Phi$-space through:
\begin{equation}
  \label{eq:HFCPtr}
\Phi\ \to\ \Phi'\ =\ e^{i\theta^a K^a}\,\Phi\; ,
\end{equation}
where $\theta^a  \in [0,2\pi)$ are  the group parameters of  the ${\rm
    SU_M(4)/Z_2}$ group.

It  is interesting  to determine  the  SO(5) generators  related to  a
particular   accidental   symmetry   that  remain   (un)broken   after
electroweak symmetry breaking.  In this way, we can find the number of
pseudo-Goldstone   bosons  predicted,   according  to   the  Goldstone
theorem.  In the  5-dimensional  bilinear ${\rm  R}^{\rm I}$-space,  a
neutral vacuum solution in  its standard basis implies that $\phi^{\sf
T}_1  i\sigma^2 \phi_2  = 0$,  i.e.~${\rm R}^4  = {\rm  R}^5 =  0$, or
equivalently ${\rm R}^\mu {\rm R}_\mu = 0$.  Alternatively, a standard
basis for writing  down a neutral vacuum solution  ${\rm R}^{\rm I}_0$
may be defined through the relation: $T^0_{\rm IJ}\, {\rm R}^{\rm J}_0
= 0$.   Consequently, an SO(5) generator $T^a$  remains unbroken after
electroweak symmetry breaking, if it satisfies the condition:
\begin{equation}
T^a_{\rm IJ}\ {\rm R}^{\rm J}_0\ =\ 0\; .
\end{equation}
By definition, the hypercharge generator $T^0$ will always be unbroken
when acting  on a neutral  vacuum solution ${\rm R}^{\rm  I}_0$.  This
should  not  be  too  surprising,   as  $T^0$  is  equivalent  to  the
electro\-magnetic  generator, given  by ${\rm  Q}_{\rm em}  = \sigma^0
\otimes \sigma^0 \otimes (\sigma^3/2) + K^0$ in the $\Phi$-space, once
we notice  that the weak isospin generator  $\sigma^0 \otimes \sigma^0
\otimes   (\sigma^3/2)$  has   no  effect   on  the   ${\rm  SU(2)_L}$
gauge-invariant 5-vector ${\rm R}^{\rm I}$.

\begin{table}[!t]
{\small 
\begin{center}
\begin{tabular}{|r|c||cc|c|l|}
\hline
 & & & & & \\[-2mm]
No & Symmetry & Generators & Discrete Group & Maximally Broken & 
Number of Pseudo-\\
 &  & $T^a \leftrightarrow K^a$ & Elements & SO(5) Generators & 
Goldstone Bosons\\[1mm]
\hline
\hline

1 & $\rm Z_2 \times O(2)$ & $T^0$ & ${\rm D}_{\rm CP1}$ & -- & 0\\

\hline

2 & $\rm (Z_2)^2\times SO(2)$ & $T^0$ & ${\rm D}_{\rm Z_2}$ & -- &  0\\  

\hline

3 & $\rm (Z_2)^3 \times O(2)$	& $T^0$ & ${\rm D}_{\rm CP2}$ & -- & 0\\

\hline

4 & $\rm O(2) \times O(2)$	& $T^3,\, T^0$ & -- & $T^3$ & 1 ~~~~($a$)\\  

\hline

5 & $\rm Z_2 \times [O(2)]^2$	& $T^2,\, T^0$ & ${\rm D}_{\rm CP1}$ &
$T^2$ & 1 ~~~~($h$)\\  

\hline  

6 & $\rm O(3) \times O(2)$	& $T^{1,2,3},\, T^0$ & -- & $T^{1,2}$ & 2 
~~~~($h,\, a$)\\ 

\hline

7 & $\rm SO(3)$                 & $T^{0,4,6}$ & -- & $T^{4,6}$ & 2 
~~~~($h^\pm$)\\

\hline

8 & $\rm Z_2 \times O(3)$ & $T^{0,4,6}$ & ${\rm D}_{\rm Z_2} \cdot
{\rm D}_{\rm CP2} $ & $T^{4,6}$ & 2 ~~~~($h^\pm$)\\

\hline

9 & $\rm (Z_2)^2 \times SO(3)$	& $T^{0,5,7}$ & ${\rm D}_{\rm CP1} \cdot
{\rm D}_{\rm CP2}$ & $T^{5,7}$ & 2  ~~~~($h^\pm$) \\

\hline

10 & $\rm O(2) \times O(3)$ & $T^3,\, T^{0,8,9}$ & -- & $T^3$ & 1 ~~~~($a$)\\

\hline

11 & $\mathrm{SO(4)}$	& $T^{0,3,4,5,6,7}$ & -- & $T^{3,5,7}$ & 3 
~~~~($a,\, h^\pm$)\\

\hline

12 & $\rm Z_2 \times O(4)$	& $T^{0,3,4,5,6,7}$ & 
${\rm D}_{\rm Z_2} \cdot
{\rm D}_{\rm CP2}$ & $T^{3,5,7}$ & 3 ~~~~($a,\, h^\pm$)\\

\hline

13 & SO(5)	& $T^{0,1,2,\dots,9}$ & -- & $T^{1,2,8,9}$ & 4 
~~~~($h,\,a,\, h^\pm$)\\

\hline

\end{tabular} 
\end{center}  }
\caption{\it  Symmetry  generators [cf.~\eqref{eq:Ka},  \eqref{eq:Ta}]
  and discrete group elements [cf.~\eqref{eq:D}] for the 13 accidental
  symmetries of the ${\rm U(1)_Y}$-invariant 2HDM potential.  For each
  symmetry, the  maximally broken  SO(5) generators compatible  with a
  neutral  vacuum are  displayed, along with the pseudo-Goldstone bosons 
(given in parentheses)  that result from the Goldstone 
theorem.\label{tab:sym}}
\end{table}

In  Table~\ref{tab:sym},  we   exhibit  the  SO(5)  (${\rm  SU_M}(4)$)
symmetry  generators $T^a$ ($K^a$)  [cf.~\eqref{eq:Ta}, \eqref{eq:Ka}]
and the  discrete group elements [cf.~\eqref{eq:D}]  generating the 13
accidental symmetries of  the ${\rm U(1)_Y}$-invariant 2HDM potential.
We also display the  maximally broken SO(5) generators compatible with
a neutral vacuum  for each symmetry, along with  the maximal number of
pseudo-Goldstone bosons  that result from the  Goldstone theorem.  The
pseudo-Goldstone bosons  associated with the maximal  breaking of each
symmetry   have  also   been  identified   in  the   last   column  of
Table~\ref{tab:sym},  using the  explicit  analytic results  presented
in~\cite{Pilaftsis:1999qt}  for the  minimization  conditions and  the
scalar mass matrices.   Thus, we find that as well  as $\rm CP1 \equiv
Z_2 \times O(2)$,  the symmetries SO(3) and $\rm  Z_2 \times O(3)$ can
maximally break  spontaneously via a CP  non-invariant vacuum.  Unlike
in the CP1 case, spontaneous breakdown of these two new symmetries may
lead to two pseudo-Goldstone bosons, i.e.~the two charged Higgs bosons
$h^\pm$.   For the symmetry  $\rm (Z_2)^2  \times SO(3)$,  the maximal
breaking   pattern  leading  to   the  two   charged  pseudo-Goldstone
bosons~$h^\pm$ is  obtained, when the restriction $\lambda_4  = - {\rm
Re}\,\lambda_5 > 0$ is taken from Table~\ref{tab:param}.

On the  other hand,  it is worth  reiterating that the  symmetry ${\rm
SO}(5)$      relates     to      the      larger     ${\rm      O}(8)$
group~\cite{Deshpande:1977rw} in the real field space, once the latter
gets  further restricted  such that  the ${\rm  SU}(2)_{\rm  L}$ gauge
canonical  form of  the $\Phi$  kinetic  term is  maintained.  In  the
5-dimensional bilinear ${\rm R^I}$-space, ${\rm SO}(5)$ can break down
to ${\rm SO}(4)$, giving rise  to four pseudo-Goldstone bosons: one of
the two CP-even Higgs bosons denoted as $h$, the CP-odd scalar $a$ and
the  two  charged  Higgs  bosons  $h^\pm$.  This  is  consistent  with
breaking pattern of ${\rm O(8) \to O(7)}$ in the $\Phi$-space, leading
to seven Goldstone bosons,  which include the three would-be Goldstone
bosons associated  with the longitudinal polarizations  of the $W^\pm$
and $Z$ bosons.  However, one  gets a different result within the $\rm
U(1)_Y$-restricted          SO(3)          bilinear          formalism
of~\cite{Nishi:2006tg,Maniatis:2006fs,Ivanov:2006yq,Ferreira:2009wh,
Maniatis:2011qu}. The  higher HF/CP symmetry~${\rm  SO(5)}$ appears as
${\rm  SO(3)}_{\rm  HF}$   in  the  $\rm  U(1)_Y$-restricted  bilinear
formalism, and  according to Table~\ref{tab:sym}  (symmetry no.~6), it
may  break down  to SO(2),  giving rise  to only  two pseudo-Goldstone
bosons.

Another illustrative  example is the symmetry ${\rm  SO}(4)$, which is
equivalent to ${\rm O(4)}\otimes{\rm O}(4)$~\cite{Deshpande:1977rw} in
the  scalar-field  space,  where  one  of  the  ${\rm  O}(4)$  factors
describes   gauge-group  transformations.    As  can   be   seen  from
Table~\ref{tab:sym},  the symmetry  ${\rm SO(4)}$  may break  to ${\rm
SO(3)}$,  giving rise  to  three pseudo-Goldstone  bosons: the  CP-odd
scalar  $a$ and  the two  charged Higgs  bosons~$h^\pm$.   Again, this
breaking  scenario cannot  be  clearly distinguished  from a  scenario
based on  ${\rm CP3  \equiv Z_2 \times  [O(2)]^2}$, which leads  to an
erroneous breaking pattern predicting only one pseudo-Goldstone boson,
within the $\rm U(1)_Y$-constrained SO(3) bilinear formalism.

It  is interesting  to  remark that  the Majorana-constrained  unitary
group ${\rm  SU_M}(4)$ in~\eqref{eq:Gmanifold} contains  the custodial
symmetry   group~${\rm  SU(2)_C}$~\cite{Sikivie:1980hm}   (for  recent
studies,    see~\cite{Grzadkowski:2010dj,Nishi:2011gc}).     In    the
$\Phi$-basis,  there  are  three  independent realizations  of  ~${\rm
SU(2)_C}$     induced    by    the     generators:    (i)~$K^{0,4,6}$;
(ii)~$K^{0,5,7}$;      (iii)~$K^{0,8,9}$.       As      stated      in
Table~\ref{tab:sym}, the HF/CP  accidental symmetries 7--13 contain at
least one  of the three  generator sets (i),  (ii) and (iii),  and are
therefore  custodial symmetric.   As  a consequence  of the  custodial
symmetry,  the $W^\pm$  and  $Z$  bosons are  degenerate  in mass  and
Veltman's $\rho$-parameter~\cite{Veltman} retains its tree-level value
$\rho = 1$,  to all orders in perturbation theory.   As happens in the
SM, however,  the ${\rm  U(1)_Y}$ hypercharge and  Yukawa interactions
violate explicitly the custodial symmetry in the 2HDM.

In  summary,   we  have   presented  the  symmetry   generators  $K^a$
in~\eqref{eq:Ka} that describe the 13 accidental symmetries~\cite{BBP}
of the ${\rm U(1)_Y}$-invariant 2HDM potential~\eqref{eq:V2HDM} in the
original scalar field space $\Phi$, by means of~\eqref{eq:HFCPtr}.  We
have  derived an  exact symmetry  relation  in~\eqref{eq:Tadef}, which
gives  the  one-to-one correspondence  between  the  ${\rm SU_M  (4)}$
generators $K^a$ in the $\Phi$-space and the SO(5) generators $T^a$ in
the  ${\rm   R}^{\rm  I}$-space.   In   Table~\ref{tab:sym},  we  have
explicitly presented  all symmetry  generators associated with  the 13
accidental symmetries, along with possible maximal breaking scenarios.
Most  importantly, we  have explicitly  demonstrated how  the bilinear
formalism  based on  the $\rm  SO(5) \subset  SO(1,5)$  symmetry group
respects  the  Goldstone theorem,  predicting  the  correct number  of
pseudo-Goldstone  bosons  after  electroweak  symmetry  breaking.   In
conclusion, the  results presented in  this note provide  firm support
for the completeness of the classification conducted in~\cite{BBP} for
the 13 accidental symmetries of the 2HDM potential.

\subsection*{Acknowledgements} This work is supported in part by
the STFC research grant, ref: ST/J000418/1.

\end{document}